\documentclass[12pt]{article}
\def\pd{\partial}
\def\bea{\begin{eqnarray*}}
\def\eea{\end{eqnarray*}}
\title{
A Coding of Real Null Four-Momenta into World-Sheet Co-ordinates} 
\author{David B. Fairlie}
\begin{document}

\maketitle

\begin{abstract}
The results of minimizing  the action for string-like systems on a simply-connected world sheet are shown to encode the Cartesian components of real null momentum four-vectors into co-ordinates on the world sheet. This identification arises consistently from different approaches to the problem.
\end{abstract}

\section{ Recapitulation}
This article is based upon an old unpublished article with David Roberts, \cite{dualmod} which dates back to circa 1972, on a model for amplitudes suggested by 
string theory, or rather the dual resonance model as it was then called. (The results in this paper are recorded in the PhD thesis of Roberts \cite{rob}).  I was so overwhelmed by the evident truth of the famous paper of Goddard, Goldstone Rebbi and Thorn quantising the bosonic string in 26 dimensions \cite{gog}, which I regard as one of the classic papers in string theory,  that I never submitted this article for publication, and after some consideration, decided not to seek publication for this version, rewritten in January of this year. However since recently there has appeared an article by Sommerfield and Thorn \cite{thorn}, the 4th section of which is closely related to the model presented in  \cite{dualmod},  it may be an appropriate time to give these ideas  an airing. Also other recent developments have  given rise to an interpretation of maximally violating helicity (MHV) amplitudes in Yang Mills theory in terms of topological string amplitudes \cite{witten},  the connection between null-four vectors and Koba-Nielsen variables which is at the heart of \cite{dualmod} may not be entirely coincidental. The intention was to construct a viable amplitude for particles living in a strictly four-dimensional space-time, and with zero mass instead of the tachyon ground state which bedevilled the bosonic string dual resonance model. One of the features of tractable models of physical processes which has come to be more appreciated in the intervening years, is that there is frequently a mis-match between what is tractable mathematically and what one should like to have; for example the potential integrability  of $N\rightarrow \infty$ supersymmetric Yang Mills as against the intractability of QCD, or the Sine Gordon model which displays  both solitons and Lorentz invariance, at the cost of working in two dimensions.
Here a feature analogous to  Self Dual Yang Mills Theory, which possesses instantons in a space of even signature is present; I have realised that the theory
presented is more mathematically compelling in a  space of signature $(2,2)$, though a Lorentzian interpretation is by no means ruled out.
This will be discussed later in relation to the work of Gross and Mende on high energy scattering.\cite{gross}
The starting point is the famous Koba-Nielsen formula, which  gives an elegant expression for the $ N$ point tree amplitude for $N$ particles with incoming momenta $p_i^\mu$ for the ground state of open strings \cite{koba} 
\begin{equation}
 { A(s,t)t\,=\, \int^\infty_{-\infty}\frac{dz_k}{dV_{abc}}\prod^N_1\theta(z_i-z_{i+1})\prod_{j>i}(z_i-z_j)^{-2\alpha'p_i.p_j}}\label{one} 
\end{equation}
\begin{equation}
 { dV_{abc}\,=\,\frac{dz_adz_bdz_c}{(z_b-z_a)(z_c-z_a)(z_a-z_c)}}.
\end{equation}
 (This integration measure is introduced as a consequence of conformal invariance; to account for the property that the real axis along which the integration is performed is invariant under transformations of the M\"{o}bius group, provided that  $\alpha'(p_i^\mu)^2=-1$ 
\begin{equation}
 {z' \mapsto \frac{az+b}{cz+d},\ \ \ ad-bc=1})\label{mob} 
\end{equation}
It has been shown that this formula arises as a contribution to string scattering from a simply connected world sheet, thanks to the properties of
conformal invariance. Another way of writing (\ref{one}) is as an exponential,
\begin{equation}
 \int_\infty^\infty\exp\left(\sum_{i<j}-2\alpha'p_i.p_j\log(z_i-z_j)\right)\frac{dz_k}{dV_{abc}}.\label{two}
\end{equation}
The exponent in the integrand may be interpreted as the (Euclidean) contribution to the action where the momenta enter the upper half plane at designated points $z_k$ which are then integrated over to give the contribution to the path integral for the amplitude arising from a simply connected world sheet.
One of the chief deficiencies in (\ref{one}) is the tachyon condition, namely that $p_i^\mu$ is light like. This requirement follows from the invariance under mappings which preserve the upper half complex plane. The radical idea behind \cite{dualmod} was to give the formula for the amplitude a different interpretation;
don't integrate, but instead determine the co-ordinates $z_i$ by minimising the integrand; this is tantamount, in the second version to using the method of steepest descents. 
The equations to be satisfied are, setting $\alpha'=1$;
\begin{equation}
 \sum_j\frac{ p_i.p_j}{(z_i-z_)}\,=\,0.\label{four}
\end{equation}
These equations may be seen to be satisfied, provided we are in a 4-dimensional space with signature $(2,2)$  with null four momenta $p_j^\mu$ and the co-ordinates $z_j$ (here on the real line), are given by 
\begin{equation}
 z_j \,=\, \frac{p_j^0+p_j^1}{p_j^2-p_j^3}\,=\,\frac{p_j^2+p_j^3}{p_j^0-p_j^1};\ \ (p_j^0)^2 +(p_j^3)^2-(p_j^1)^2-(p_j^2)^2=0.\label{conn}
\end{equation}
This works because
\begin{eqnarray}
 (z_i-z_j) &=& \frac{p_i.p_j+p_i^0p_j^1-p_i^1p_j^0+p_i^3p_j^2-p_i^2p_j^3}{(p_i^2-p_i^3)(p_j^0-p_j^1)}\nonumber\\
 (z_i-z_j) &=& \frac{-p_i.p_j+p_i^0p_j^1-p_i^1p_j^0-p_i^3p_j^2-p_i^2p_j^3}{(p_i^0-p_i^1)(p_j^2-p_j^3)}\nonumber
\end{eqnarray}
The second equation is obtained by using the alternative expression of $(z_i,\ z_j)$. Subtracting, and rationalising, we have
\begin{equation}
 (p_i^2-p_i^3)(p_j^0-p_j^1)-(p_i^0-p_i^1)(p_j^2-p_j^3)\,=\, \frac{ 2p_i.p_j}{(z_i-z_j)}. \label{five}
\end{equation}
Summing over all particle positions $z_j$ except $z_j=z_i$ and invoking the conservation of momentum, $\sum p_j^\mu\, =\,0$, we see that (\ref{four}) is satisfied. If instead of the real line, the integration in (\ref{one}) is performed over the boundary of the unit disc, the points  on the boundary where the momenta enter may be parametrised by
\begin{equation}
 z_j \,=\, \frac{p_j^0+ip_j^3}{p_j^1+ip_j^2}\,=\,\frac{p_j^1-ip_j^2}{p_j^0-p_j^3};\ \ (p_j^0)^2 +(p_j^3)^2-(p_j^1)^2-(p_j^2)^2=0.\label{disc}
\end{equation}
There is a M\"obius transformation (\ref{mob}) which connects the two representations, for the plane and the disc;
\begin{equation}
 z_{\rm disc} = \frac{i+z_{\rm plane}}{i -z_{\rm plane} }\label{mob2}
\end{equation}
Indeed complex M\"obius transformations on $z_i$ are equivalent to $SU(2,2)$ transformations on $p_i^\mu$.
\section{Alternative Approach}
Consider a two dimensional surface embedded in a four dimensional space and take as parametric representation of the surface the four vector $X_\mu(\sigma,\tau)$
where $\sigma$ and $\tau$ are intrinsic  co-ordinates on the surface with metric
\[ ds^2 \,=\, Ed\sigma^2 +2Fd\sigma d\tau +G d\tau^2,\]
where
\[ E=\left(\frac{\pd X_\mu}{\pd\sigma}\right)^2,\ \ F=\left(\frac{\pd X_\mu}{\pd\sigma }\right)\left(\frac{\pd  X_\mu}{\pd \tau}\right)^2;\ \ G=\left(\frac{\pd X_\mu}{\pd\tau}\right)^2.\]\cite{dualmod}
The Nambu-Goto Lagrangian describing the dynamics of the field $X_\mu(\sigma,\tau)$ is a measure of the area of the world sheet and is the reparametrisation invariant form
\begin{equation}
 {\cal L} \,=\, \alpha'\int\int\ \sqrt{EG-F^2}d\sigma d\tau.\label{lag}
\end{equation}
On the other hand it is well known that there exists a transformation to a co-ordinate system of so-called isometric co-ordinates in which the Lagrangian takes  
the simple quadratic form
\begin{equation}
 {\cal L'}\, =\, \int\int \left(\left(\frac{\pd X_\mu}{\partial\sigma}\right)^2+\left(\frac{\pd X_\mu }{\pd\tau}\right)^2\right)d\sigma d\tau.\label{lag2}
\end{equation}
which is invariant only under the subset of reparametrisations  of the variables $(\sigma,\ \tau)$ which are conformal, i.e., those transformations which satisfy the Cauchy-Riemann equations.It is well known that in the co-ordinate system where $\sigma$ and $\tau$ are isometric parameters defined by $E\,=\,G;\ \ F\,=\,0$ In this frame the Euler equation minimising (\ref{lag}) becomes linear and is just
\begin{equation}
 \nabla^2X_\mu\,=\,0.\label{laplace}
\end{equation}
The conditions for an isometric co-ordinate system may be written in the form due to Weierstrass;
\[\left (\frac{\pd \zeta_\mu}{\pd z}\right)^2\, =\, E-G +2iF \,=\,0,\]
where $X_\mu$ is the real part of $\zeta_\mu$ in view of the fact that (\ref{laplace}) is satisfied provided $\zeta $ is an analytic function of
$z\,=\, \sigma +i\tau$. The Weierstrass condition shows that conformal mappings of co-ordinate systems preserve the isometric property.
We can make a link with the Virasoro conditions for closed strings \cite{vir1}  by noting that this is in fact the gauge condition of the model; writing
\begin{equation}
 \left(\frac{\pd\zeta_\mu}{\pd z}\right)^2 \,=\, \sum^{\infty}_{-\infty}L_nz^n\,=\,0.\label{vir1}
\end{equation}
This is too stringent to demand as an operator equation.  Instead we require that the matrix elements of (\ref{vir1}) should vanish for all z; i.e.
\[ <\psi^\dag | L_n|\psi>\,=\,0, \forall n\geq 0\]
This is satisfied provided $L_n\,=L_n^\dag \,=\, 0$ these conditions are the familiar Virasoro conditions for closed strings with zero mass ground states.
A typical solution of (\ref{laplace}) with a finite number of singularities is given by
\begin{equation}
 \zeta^\mu \,=\, \sum^{i=n}_{i=1} p^\mu_i\log(z-z_i) \label{solln}
\end{equation}
Applying the Weirestrass condition, we have
\begin{equation}
 \sum_{i,j}\frac{p_i.p_j}{(z-z_i)(z-z_j)}\,=\, \sum_{i,j}{p_i.p_j}\left (\frac{1}{(z-z_i)(z_j-z_i)}-\frac{1}{(z-z_j)(z_j-z_i)}\right)=0\label{condd}
\end{equation}
This has to be true for all $z$, which evidently requires that $\sum_{i,j} p_i.p_j\,=\,0$ and, with conservation of four momentum, the same conditions
$\displaystyle{\sum_j\frac{p_i.p_j}{(z_i-z_j)} \,=\,0}$ as before.
In the case of the four point function the solution of these conditions, (\ref{four}) may be readily solved in terms of the cross-ratio $\lambda$ to give
\begin{equation}
 \lambda\,=\, \frac{(z_i-z_2)(z_3-z_4)}{(z_1-z_3)(z_4-z_2}\,=\, \frac{p_1.p_2}{p_1.p_3}\,=\, \frac{s}{t}\label{cross}
\end{equation}
where 
\[s\,=\,(p^\mu_1+p^\mu_2),\ t\,=\,(p^\mu_1+p^\mu_3),\ u\,=\,(p^\mu_1+p^\mu_4);\ \  s+t+u\,=\,0\]
This result, in terms of the cross ratio is independent of the metric, so also works in a Lorentz metric with signature $(3,1)$. The resulting amplitude $A(s,t,u)$ with $s+t+u\,=\,0$ may be evaluated to give
\begin{equation}
 A(s,t,u)\,=\, (-s)^{-\alpha' s}(-t)^{-\alpha' t}(-u)^{\alpha' u}.\label{amp}
\end{equation}
As $s\rightarrow\infty$ at fixed t $A(s,t,u)\rightarrow  t^{-\alpha' s}s^{-\alpha' t}$ i,e., it exhibits Regge asymptotic behaviour. The subject of asymptotic behaviour of high energy string amplitudes was examined to all orders sometime afterwards by Gross and Mende \cite{gross} who found the same connection (\ref{cross}) between the cross-ratio and the Mandelstam variables.
\section{Lorentz signature}
As has been remarked, the minimisation condition in the case of the four-point function may be solved in terms of cross ratios. This suggests that the conditions
may be solved directly in terms of the variables $z_j$ whatever the metric. This is indeed the case; for real four momenta and, the solution may be expressed as
\begin{equation}
 z_j \,=\, \frac{p_j^0+p_j^3}{p_j^1-ip_j^2}\,=\,\frac{p_j^i+ip_j^2}{p_j^0-p_j^3};\ \ (p_j^0)^2 -(p_j^1)^2-(p_j^2)^2-(p_j^3)^2=0.\label{conn2} 
\end{equation}
The difference is that in the case of signature ${2,2}$ the four momenta may be parametrised as $p_j^0=r\cosh(\theta_j),\ p_j^1=r\sinh(\theta_j),\ p_j^2=r\cosh(\phi_j),\ p_j^3=r\sinh(\phi_j)$, which imply
\begin{equation}
 z_j\,=\, \exp(\theta_j+\phi_j)\label{real}
\end{equation}
 so the variables lie on the real. line Alternatively a trigonometric parametrisation may be employed, in which case $z_j\,=\, \exp (i\theta_j+i\phi_j)$ 
However in the case of signature ${1,3}$ the parametrisation is mixed;$p_j^0=r\cosh(\theta_j),\ p_j^3=r\sinh(\theta_j),\ p_j^1=r\cos(\phi_j),\ p_j^2=r\sinh(\phi_j)$, which imply $z_j\,=\, i\exp (\theta_j+i\phi_j)$, so there is no obvious integration contour for (\ref{one}).
\section{Minimal surface interpretation}
Further insight may be gained by a parametrisation of minimal surfaces embedded in four dimensional Euclidean space, originally due to Eisenhart, \cite{eis}
but rediscovered by  Shaw \cite{shaw} and quoted in \cite{dualmod} and \cite{cor1}.
It is given by
\begin{eqnarray*}
 X^0 &=& {\it Re}(f(z)-zf'(z)+g')\\
 X^3 &=& {\it Im}(f(z)-zf'(z)-g')\\
  X^1 &=& {\it Re}(g(z)-zg'(z)+f')\\
  X^2 &=& {\it Im}(zg'(z)-g(z)+f')\\
\end{eqnarray*}
Suppose we seek a parametrisation where $X^\mu +a^\mu$  is the real part of  $\zeta^\mu \,=\,\sum_i  p_i^\mu G_i(z)$ and $a^\mu$ is an arbitrary origin. Then, thanks to the linearity of the above equations, we can split $f(z)$ and $g(z)$ into sums of independent components; i.e. $f(z)\,=\,\sum f_i(z);\ \  g(z)\,=\,\sum g_i(z)$ and
deduce, up to shifts of origin,
\begin{eqnarray*}
(p_i^0+ip_i^3)G_i(z)&=& 2g_i'(z)); \ \ (p_i^1-ip_i^2)G_i(z) \,=\, 2(g_i(z)-zg_i'(z))\\
(p_i^1+ip_i^2)G_i(z)&=& 2f_i'(z)); \ \ (p_i^0-ip_i^3)G_i(z) \,=\, 2(f_i(z)-zf_i'(z))\\
\end{eqnarray*}
These equations may be easily solved to give
\begin{eqnarray*}
 2g_i(z)&=& c(z-z_i)\log(z-z_i)-z,\\
 2f_i(z)&=& \frac{c}{z_i}\left((z-z_i)\log(z-z_i)-z\right);\\
 G_i& =& c\log(z-z_i).
\end{eqnarray*}
where $c$ is a constant.
\begin{equation}
 z_i = \frac{p_i^0+ip_i^3}{p_i^1-ip_i^2}\,=\,\frac{p_i^1+ip_i^2}{p_i^0-ip_i^3}\label{consist}
\end{equation}
The last relation follows from consistency. These are the same relations between the co-ordinates $z_i$ and the momenta found before.
If this constant is pure imaginary and the real parametrisation (\ref{five}) is employed, so the $z_i$ lie on the real axis
\begin{equation}
X^\mu = \sum {\it Re}p^\mu_i\log(z-z_i) = \sum \pi  p^\mu_i\Theta(z-z_i)/ / {\rm for}/  z\ {\rm  on\ the\ real\ axis}
\end{equation}
As $z$ moves from $ +\infty$ to $-\infty$, $X^\mu$ jumps by $\pi p^\mu_i$, so the skew polygon formed by the partial sums of momenta (closed on account of momentum conservation) is mapped into intervals on the real line.
\section{Conclusion}
The principal message of this article is to draw attention to the link between the Cartesian components of  real null four-momenta in four dimensional flat space and complex variables on a simply connected world sheet, associated with a minimal surface, or a form of string evolution. The set of four-momenta are also required to sum to zero; i.e momentum is conserved in the system. Various aspects leading to this identification are explored: The minimisation of the Koba-Nielsen integrand; the consequence of  the  Weierstrass condition  upon a  linear combination of elementary solutions to the free equations of motion, to guarantee a minimal surface solution and the direct determination of  this class of minimal surface solution from the Eisenhart  parametrisation  are all shown to entail the same identification of a complex variable in terms of the components of a null four momentum. In a space of even signature $(2,\ 2)$; in one representation the complex variables  lie on the real line; in another on a circle; in the case of odd signature, (Lorentz  metric), there is no specific curve on which the variables lie.  $SL(2,C)$  transformations of the complex variable implement
homogeneous Lorentz  transformations upon the momentum.

In our original article, as is standard practice, the  optimistic anticipation of further development of these ideas was raised, but it must be admitted that neither author has been able to add anything substantially new in the intervening 35 years! However, as T.S. Eliot has said, `A poem may have meanings which are hidden from its author’. The recent paper of Sommerfield and Thorn \cite{thorn} extends their ideas to AdS space-time, and the picture of world sheets bounded by a closed polygon of null lines which is presented therein and is also contained in \cite{alday} is essentially the same as that in section 4 of the present article
In addition, the treatment of high energy string amplitudes by Gross and Mende \cite{gross} extends some aspects of this analysis to multiply connected world sheets.

In this spirit this revised and rewritten version of \cite{dualmod} is offered in the hope that some deeper connection between momentum space and the world sheet will be discovered.


\begin{thebibliography}{**} 
\bibitem{dualmod} D.B. Fairlie and D.E. Roberts, Dual Models without Tachyons - a new approach, unpublished ca. 1972.
\bibitem{rob} D.E. Roberts, Mathematical Structure of Dual Amplitudes, PhD thesis (unpublished), Durham University Library (1972), chapter IV.
\bibitem{thorn} C.B. Thorn and C.M. Sommerfield,  Classical Worldsheets for String Scattering on Flat and AdS Spacetime {\bf arXiv:hep-th/88} (2008)
\bibitem{witten} E. Witten, Perturbative gauge theory as a string theory in twistor space, {\it Comm.Math. Phys.} {\bf 252} (2004) 189-258.
\bibitem{koba} Z.Koba and H.B. Nielsen, {\it Nuclear Physics } {\bf B10} (1969) 633.
\bibitem{gog} P.Goddard, J. Goldstone, C. Rebbi and C. Thorn {\it Nuclear Pysics B} {\bf 56} (1973) 109.
\bibitem{fai1} D.B.Fairlie and H.B. Nielsen, An Analogue Model for K.S.V. Theory,
{\it Nuclear Physics}~{\bf B20} (1970) 637- 651.
\bibitem{gross} D.J. Gross and P. F. Mende, The High-Energy Behaviour of String Theory, {\it Phys. Lett.} {\bf B197} (1987) 129.
\bibitem{vir1} M. A. Virasoro{\it Il Nuovo Cimento} {\bf 64 A} {1969} 811-840, Subsidiary conditions and ghosts in dual-resonance models, {\it Phys.Rev.} {\bf D1} (1970) 2933-2936.
\bibitem{eis} L.P. Eisenhart, {\it Am. J. Math},{\bf 49}, (1912) 769.
\bibitem{shaw} W.T. Shaw, {\it Class. \& Quantum Gravity} {\bf 2} (1985) {\bf L}113.
\bibitem{cor1} D.B.Fairlie and C.A. Manogue, Lorentz Invariance and the Composite String,
{\it Physical Review }~{\bf D34} (1986) 1832-1834.
\bibitem{alday} L.F. alday and J Maldacena, {\it JHEP} {\bf 0711} (2007) {\bf aXiv:hepth/0710.1060}.
\end{thebibliography}
\end{document}